\documentclass[conference,10pt,letterpaper]{IEEEtran}
\usepackage[utf8]{inputenc}   
\usepackage[english]{babel}
\usepackage{blindtext}
\usepackage{textcomp}
\usepackage{siunitx}

\usepackage{cite}
\ifCLASSINFOpdf
  \usepackage[pdftex]{graphicx}
  \DeclareGraphicsExtensions{.pdf,.jpeg,.png}
\else
  \usepackage[dvips]{graphicx}
  \DeclareGraphicsExtensions{.eps}
\fi
\usepackage[cmex10]{amsmath}
\usepackage{amssymb}
\usepackage{algorithmicx}
\usepackage{algpseudocode}
\usepackage{tikz}
\usetikzlibrary{shapes.geometric, arrows}
\tikzstyle{arrow} = [thick,->,>=stealth]

\begin{document}
\title{Topology Management, Multi-Path Routing, and Link Scheduling for mmW WMN Backhaul}

\author{\IEEEauthorblockN{Kari Seppänen}
\IEEEauthorblockA{VTT Technical Research Centre of Finland\\
Espoo, Finland\\
Email: Firstname.Lastname@vtt.fi}
\and
\IEEEauthorblockN{Pekka J. Wainio}
\IEEEauthorblockA{Nokia Bell Labs\\
Espoo, Finland\\
Email: Firstname.Lastname@nokia-bell-labs.com}
}

\maketitle

\begin{abstract}
Mobile backhaul system based on a wireless mesh network using
point-to-point millimetre wave links is a promising solution for dense
5G small cell deployments. While mmW radio technology can provide
the sufficient capacity, the management of transport delays over
multiple wireless hops is challenging especially if TDD backhaul
radios are used. Earlier, we have proposed the Self-Optimizing WMN
concept and presented routing and link scheduling principles that can
be used for backhaul nodes with single radio unit. In this paper, we are
extending the concept to support backhaul nodes that can have multiple
radio units with own beam steering antennas covering non-overlapping
sectors. The proposed system is based on dividing the task in separate
phases. In the first phase, an active network topology is created by
selecting a suitable subset of all available links. In the next step,
the routing information and transmission sets are generated. Finally,
the link schedule is optimized by finding an optimal ordering of
transmission sets. In this paper, we are proposing a feedback loop from transmission set generation to topology management. We show that this feedback loop removes efficiently "troublesome" links from the active topology making it easier to find optimal link schedules.
\end{abstract}

%
\IEEEpeerreviewmaketitle

\section{Introduction}

Network densification~\cite{Bhushan2014}, that is, augmenting marco
cell capacity with a large number of small cell base stations (SB), is
considered to be one of the viable solutions to meet 5G capacity
demand. New types of installation locations for SBs (light poles, bus
stops) create a new challenge due to lack of existing optical fiber
connectivity. Thus, wireless backhaul (BH)
utilizing high frequency millimeter wave (mmW) connections is proposed
~\cite{Dehos2014,Shariat2015}. These connections utilize high gain
beam forming antennas that support high bandwidth, high spatial reuse factor, and thus
high areal capacity.

The mmW BH networks rely mostly on line-of-sight (LOS) links and thus
attenuation and blocking can cause serious problems to
reliability. Moreover, direct connectivity between SB site and BH
gateway (GW) node with fiber optic connectivity cannot be always
guaranteed, which makes it necessary to have multihop wireless
paths. The multihop connectivity makes the reliability problem even
more pronounced. To overcome these problems, partial mesh topology or
wireless mesh network (WMN) can be used to provide multi-path
connections and thus much improved fault tolerance.

We have proposed a wireless mesh network WMN based mmW BH
system~\cite{Wainio2016} that utilizes a set of routing, link
scheduling and queueing algorithms to provide high reliability and
low delays over BH segment~\cite{Seppanen2016b,Kilpi2017,Seppanen2016a}. 
The proposed system was based on a WMN node (WN) architecture with
single mmW radio unit (RU). In this paper, we extend that concept by
introducing support for multi-RU WNs with beam steering sector
antennas. The new WN architecture brings new constraint to link
scheduling algoritm as all RUs in one WN can either transmit or
receive at the same time. Furthermore, we have modified the 
routing and link scheduling process by having tighter integration
between subsequent steps and a feedback loop between topology
management and link scheduling.


\section{WMN based backhaul}
\label{sec:arch}

Our Self-optimizing WMN (SWMN) concept is based on using high capacity E--band
mmW radio links with high gain beam steering
antennas~\cite{Wainio2016}. We are assuming that interference is not a
limiting factor for the total throughput~\cite{7110547}. Thus, our
link scheduling algorithm is focused on minimizing worst case
transport delays~\cite{Kilpi2017}. The fault tolerance is provided by
multipath routing that is based on Multiple Disjoint Spanning Trees
(MDST)~\cite{Seppanen2016b}. Those spanning trees are constructed so
that the number of node disjoint routes to GW nodes are close to
maximum. Furthermore, the unfairness problem with multihop wireless
connections is solved by a specific fair queueing
mechanism~\cite{Seppanen2016a}.


\subsection{System architecture}

In the earlier work, we have assumed that each WMN Node (WN) is
composed of single TDD BH radio unit with a beam steering
antenna system that can cover full \ang{360}. Now we are considering a
node architecture, where each WN has multiple BH radios that
cover non-overlapping sectors. All radio units are still operating in TDD
mode and in the same frequency.


There are also some assumptions about physical system configuration
and construction. The components of single WN unit should be tightly
integrated into a compact rugged casing, which means that the system
dimensions are quite close to the antenna dimensions. Furthermore, the
system should be cost optimized meaning that there cannot be any
expensive components to mitigate internal interference between
RUs.

\begin{figure}[!t]
\centering
\includegraphics[width=80mm]{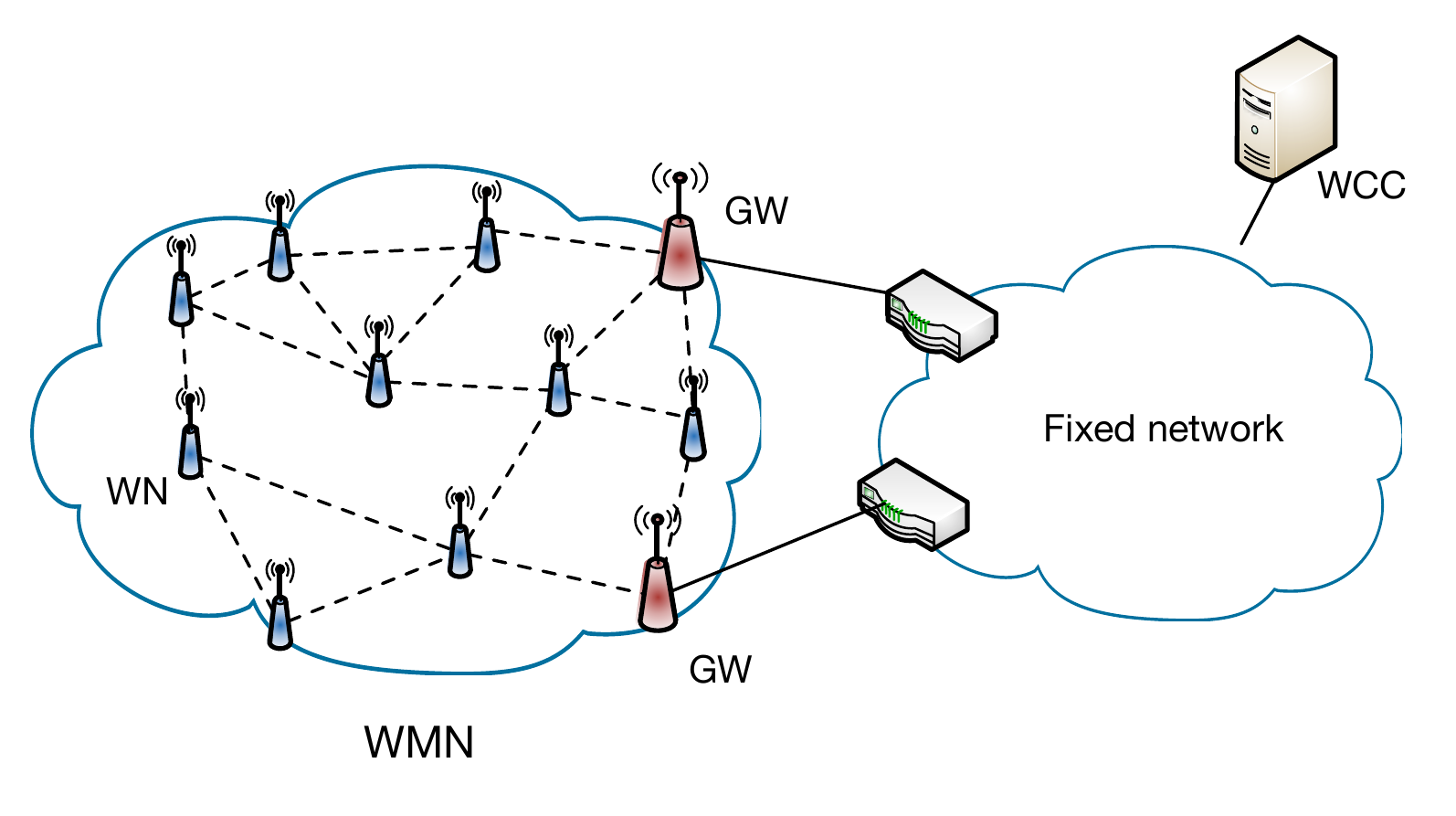}
\caption{SWMN architecture.}
\label{fig:arch}
\end{figure}

Otherwise, the network architecture is the same (see
figure~\ref{fig:arch}). All WNs perform periodically neighbor
search and they also constantly monitor the link quality. Any changes
to network topology or link performance are reported to WMN
Centralized Controller (WCC) via gateway nodes (GW). WCC is
responsible of computing a new network configuration consisting of the selected active links,
routing trees and link schedule when necessary. The routing trees
provide multiple disjoint routes for each WN to one or more GWs and
each WN is responsible of selecting dynamically the most suitable route for each
traffic flow considering detected link faults and possible congestion. This means that
while the network configuration is centrally computed and more or less
semi-static, the network can react, e.g., link failures at real-time.

\subsection{System requirements and limitations}

The proposed node architecture implies one critical constraint. Due to near-field interference between RUs in the node one RU can not transmit when any of the physically adjacent RUs is receiving and vice versa. To overcome this we have selected architectural approach where all
the RUs in one node can either send or receive at the same
time. This means that the link scheduling algorithm
should also consider link directionality and whether the link
end-points are in suitable mode (TX/RX).

As in our earlier work, we assume that end-to-end delay is the most
important optimization criteria in WMN BH. Due to natural delays
caused by TDD links, we have determined that to meet mobile BH
delay requirements, it is necessary to have short cyclic (semi)permanent
link schedules.
Thus, we take a similar
approach with this new type of node architecture and try to maximize
the network capacity by maximizing the number sectors (instead of
links) that can be active simultaneously.

\section{Routing and link scheduling algorithm}
\label{sec:alg}

\begin{figure}[!t]
\centering
\includegraphics[width=70mm]{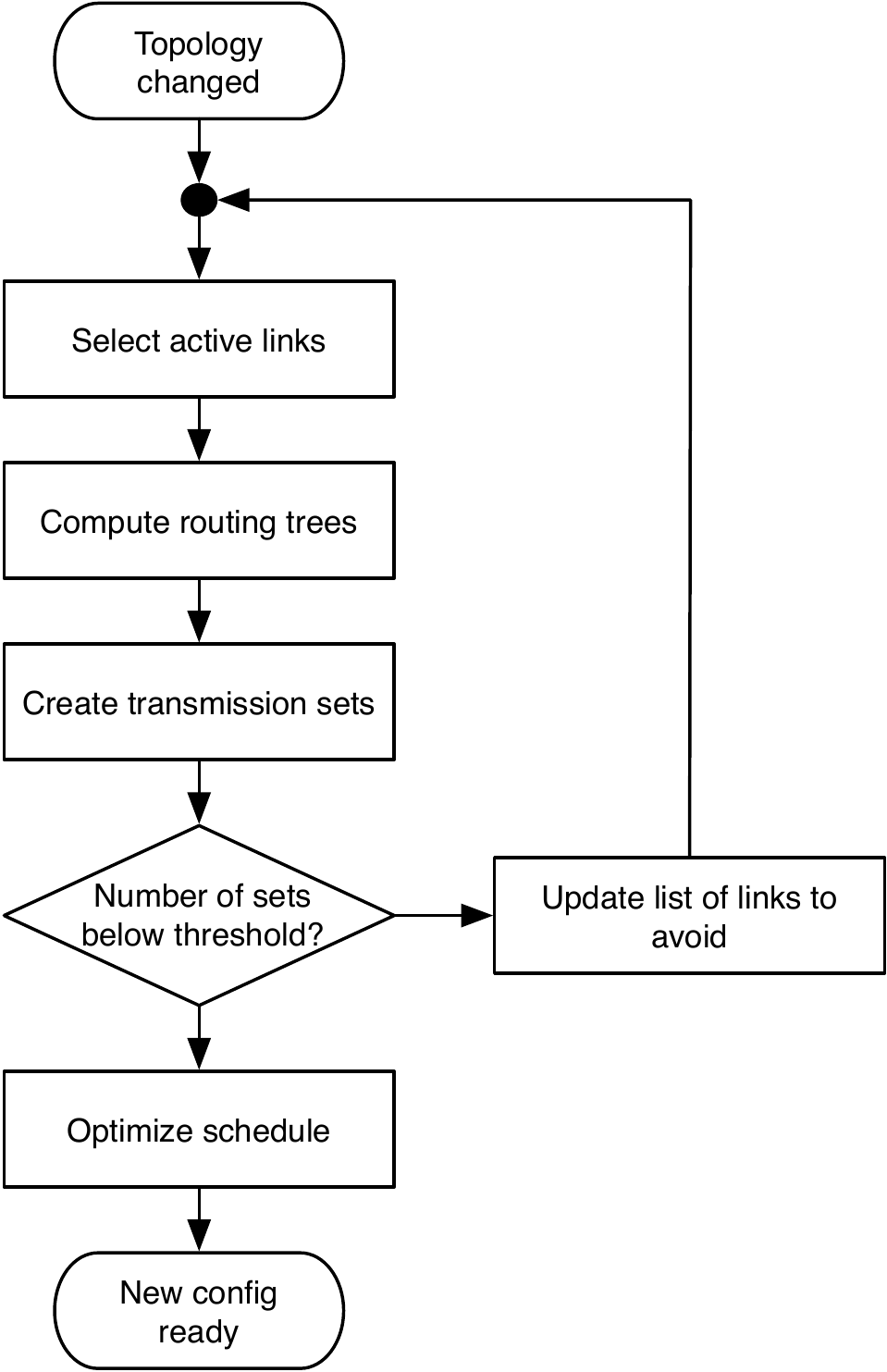}
\caption{Routing and link scheduling process.}
\label{fig:flow}
\end{figure}

The proposed new routing and link scheduling algoritm has four main
phases as shown in figure~\ref{fig:flow}:
\begin{enumerate}
\item In ``Select active links'' phase, a subset of all available links
  are chosen to be included in routing and link scheduling (so called
  ``active topology'')
\item In ``Compute routing trees'' phase, a set of spanning trees are
  computed to be used for routing and a primary path is allocated to
  each node.
\item In ``Create transmission sets'' phase, a set of maximal
  transmission sets $TSS$ is created to be used in link scheduling.
\item In ``Optimize schedule'' phase, an optimal ordering for
  transmission sets is sought.
\end{enumerate}
The process has one iterative feed-back loop between last two
phases. If the chosen active topology requires a large $TSS$ to cover
all active links, the optimization of link schedules would be
computationally challenging. Thus, if  $|TSS|$ exceeds certain
treshold, the links that causes most troubles in $TSS$ creation, are
added to ``links to avoid'' list and the whole process is started
again from the 1st phase. When the overall performance is considered, this
is quite justified as the last phase is the only computationally
challenging phase.

\subsection{Active link selection}

Our primary target is to minimize the maximum end-to-end delay. There
are two major factors that define the range of worst case delays: path length and
schedule length. While the link schedule determines the actual delays
for each node, those two factors set the boundaries to link schedule
optimization. Thus, it is important to try to choose such active
topology that would lead to optimal schedule and path length combination.

\subsubsection{Optimal fan-out}
We have found out that both of these factors depend on fan-out $k$,
that is, number of active links per sector~\cite{Seppanen2019a}. If we
assume that all active links are carrying traffic and thus need time
in the schedule, the link schedule length $|s|$ is related directly to
$k$ by $|s| \sim k$. Furthermore, if we assume that traffic is mainly
between normal nodes and gateway, all paths can be modeled with a
tree. In this case, the maximum path length $p$ in a network with $N$
nodes is related to $\max |p| \sim \log_k N.$
By combining these two relations, it can be argued that maximum delay $d$
is approximately bounded by
\[\max d \sim k \log_k N.\]

Actually, the situation is a bit more complicated, especially if we take
partial mesh topology and spatial constraints in to account (see~\cite{Seppanen2019a}
 for detailed analysis). However, this relation holds quite well and it
 indicates that we should use a quite small number of active links
 (2--3) per sector.

 \subsubsection{Link selection algorithm}
 The active link selection algorithm works as follows:
 \begin{enumerate}
 \item All links are given weight value according to traffic estimates
   or actual traffic statistics
 \item All GW nodes are set as starting points by inserting them to
   ``current nodes'' list
 \item For each node in the list
   \begin{enumerate}
   \item For each sector in the node
     \begin{enumerate}
        \item Until max number of links have been added for the sector, add new link to topology if the other end of the link is connected to a RU with less than max no. of links
     \end{enumerate}
   \item Make a new ``current nodes'' list from the new nodes that
     were just added to active topology and repeat 3) until list is empty
   \end{enumerate}
 \end{enumerate}


 If there are nodes that are left without any active links, the
 procedure can be repeated with increased maximum number of active
 links.
 
 This is a greedy algorithm that operates without the global
 view. However, it seems to produce quite reliably such active network
 topologies that provide sufficient resiliency as well as low worst
 case delays. This means that the resulting topology is such that the
 routing algorithm can find multiple node disjoint routes and the link
 scheduling algorithm can construct a short cyclic link schedule where
 all primary routes meet the delay constraint.

 \subsubsection{Constrained topologies}
 The link selection algorithm described above contains only one
 constraint, i.e., maximum number of links per sector. This does not
 limit the network topology besides setting an upper limit to node
 degree. As a consequence, there will be odd cycles that are know to
 be more difficult for link scheduling (ref.\ link coloring) than even
 cycles. We can force the network topology to contain only even cycles
 by ensuring that the network graph is bipartite. From the properties
 of bipartite graphs, it is known that a bipartite graph has an edge
 coloring using number of colors that is equal to maximum node degree~\cite{bollobas1998modern}. This
 should make finding an optimal link schedule much easier.

 One way to restrict network topology to bipartite graphs is to color
 vertices with two colors and to allow links only between differently
 colored vertices. The basic link selection algorithm can be modified
 to do vertex coloring incrementally as new nodes are being
 added to active topology. This requires two modifications:
 \begin{enumerate}
 \item Link availability depends not only on maximum number of links
   per sector but also node colors --- link can be selected only if
   end-points are of different color, or the other end is colorless
 \item If a link with a colorless end-point is selected, the colorless
   end-point is colored with the other color than the already colored end-point
 \end{enumerate}
 In one sense, the selected WMN architecture is quite well suited for such coloring
 as each node can either transmit or receive at the same time. Thus,
 two-color vertex coloring can serve as a TX/RX schedule for WNs.

 While it might be easier to find a link schedule for bipartite
 network topology, the overall impact on worst case delays, resilience, and network
 capacity is not so clear. This constraint may lead to sparser network
 and longer paths and thus it is necessary to evaluate both
 unrestricted and bipartite network cases.

\subsection{Routing trees}

We are using a multipath routing algorithm that is based on the
earlier work presented in~\cite{Seppanen2016b}. Routing is based on
multiple spanning trees (MDST) that
are constructed so that there are multiple node disjoint routes
between each node and the GW nodes. MDST configuration is computed at
WCC and distributed to WNs. Each WN uses MDST information to construct
local forwarding tables that are used to make the route selection
for each packet flow. In this way, the system can react to link
failures immediately.

MDST computation uses a greedy algorithm that is quite similar to link
selection algorithm. The process is composed of three phases:
\begin{enumerate}
\item The links (active links selected in the previous step) are
  weighted by computing shortest path from each WN to all GWs and
  allocating the estimated traffic to shortest path links.
\item Stems are constructed for each ST
  \begin{itemize}
  \item All STs are rooted at GW node
  \item Each link connected to GW serves as a starting point for a ST
  \item New links are added to stems starting from highest weight
  \item Each WN can belong to only one stem and the process end when
    all WNs belong to some stem
  \end{itemize}
\item Each stem is expanded to a full ST by adding new links according
  to weight values.
\end{enumerate}
While also this algorithm operates without global view, it tends to
produce sufficient amount $(\geq 2)$ of node disjoint routes. The main difference
to the original algorithm is that all stems are grown at the same time
instead of one GW at the time. This modification allows to algorithm
to handle WMNs with high GW-to-WN ratios.

After MDST routing trees are generated, so called primary path is
allocated to each plain WN. The primary path is usually the lowest
cost path to a GW node but if there are multiple (almost) equal cost
paths then some coarse load balancing between GWs and routing trees can
be done. These primary paths are given as an input to link scheduling
process.


\subsection{Link scheduling}

Link scheduling is based on earlier algorithm presented
in~\cite{Kilpi2017}. We are using short cyclic link schedule that is
computed by WCC. So called transmission set defines which links can be
active in a time-slot. The link schedule spans over multiple
time-slots and contains different transmission sets that cover all the
active links. The algorithm has two phases:
\begin{enumerate}
\item Construction of maximal transmission sets
\item Link schedule optimization by finding an optimal ordering for
  transmission sets
\end{enumerate}
The new modification to this algorithm is that it is now clearly
separated  in to these two phases and a feedback loop to link
selection step is added between them. Furthermore, the system
constraint where all RUs in one node can either transmit or receive at
a time is taken into account.

\subsubsection{Construction of transmission sets}

Transmission sets ($TS$) are generated using the \textsc{GreedyTwice} algorithm
from~\cite{Kilpi2017}. The algorithm is based on selecting new,
non-conflicting links to $TS$ according to weigh values and favouring
the links that are not yet included in any $TS$. The link weights are
computed by counting how many primary paths cross a specific link. In
this way, the most important links are given more importance in link
selection procedure.

\textsc{GreedyTwice} requires few modifications to handle the TX/RX
constraint. First of all, the links must be directional so that the TX
and RX end-points can be defined in the link schedule. This was not
necessary for 1 RU nodes that could use one time-slot for
bi-directional communications. Furthermore, for each time-slot, the
algorithm must define, which nodes are in TX mode and which in RX
mode. Then the link selection must ensure that the end-points of the
link are in correct TX/RX mode.

This modification is done by populating the link list, that is used by
the algorithm, with directional links. In the original version there
was just undirected link $(u,v)$ and now we are having two directed
links $(u,v)$ and $(v,u)$. The link selection has an additional step,
i.e., for link $(u,v)$:
\begin{itemize}
\item If $u$ is in RX mode or $v$ is in TX mode, link is rejected
\item If TX/RX mode for $u$ is not defined, the mode is set to TX 
\item If TX/RX mode for $v$ is not defined, the mode is set to RX 
\end{itemize}
For each TS construction round, TX/RX mode for all nodes are cleared
and thus the link scheduling step also defines the TX/RX schedule for the nodes.

In the original algorithm, the new links are selected one by one. The
same can be done with the multi-RU WMN but optionally we can try to
select also link for all sectors for a node. This is done by an
additional step to link selection: if a link is selected and it has an
end-point without TX/RX mode set, the correct mode is set and one link
for each free sector is selected if possible.


\subsubsection{Feedback loop}

The feedback loop is based on finding out, which links are causing a large number of
transmission sets to be created to cover all links. As it happens, the \textsc{GreedyTwice}
algorithm can used to perform this task by minor modifications. It
turns out that the links that are last to be included to a transmission
set, are the most likely ``troublesome links''. Because they are
selected last, it is likely that they have low weight value (i.e.,
they have lesser importance) and they are in conflict with many other
links.

The feedback loop is activated at the end of transmission set
construction. If the number of transmission sets is larger than a
threshold value (e.g., 8), the links that were not included in any
transmission set before the last round of \textsc{GreedyTwice} are added to
``links to avoid'' list. Then the whole routing and link scheduling
process is restarted from ``Select active links'' step with this list
as an input.

This is repeated until the number of transmission sets is within
desired limits. In some pathological cases, adding links to ``avoid''
list may lead to a loss of network wide connectivity. In such cases,
longer link schedules have to be accepted.

\subsubsection{Link schedule optimization}

The link schedule optimization procedure for multi-RU architecture is very similar to the previous one RU case. If the schedule length is short (e.g., $\leq 8$), a
brute force method is used, i.e., all permutations are evaluated. For
longer schedules, random shuffling and simulated annealing are
applied. However, thanks to the new feedback loop, long link schedules
are very rare.

While the link schedule is optimized only for primary paths, it turns
out that other high order paths usually have low delays. The reason
for this is that such paths contain only one unoptimized hop (the
first one) while the rest of the path is an optimized primary path for
neighboring node.

\section{Tests}
\label{sec:tests}

Test were conducted using actual WCC code that is written in
Python. The code uses mainly standard Python libraries but
NetworkX~\cite{SciPyProceedings_11} is used to provide some basic data
structures for network topology operations and PyMongo for accessing network
topology and configuration database at WCC.

We are evaluating four different strategies denoted by  \textbf{BS}:
bipartite topology, select links one by one,  \textbf{BA}:
bipartite topology, select all available links,  \textbf{FS}:
free form topology, select links one by one,  \textbf{FA}:
bipartite topology, select all available links.

\subsection{Topology model}

We are using 2.5D model where location of each node is defined by
three coordinates but all antenna planes must be perfectly
vertical. The topology model is trying to emulate normal urban environment.


As there are virtually no existing reference topologies with fixed
node locations, we are using a topology generator to make sample
network deployments that should have the typical dense urban are
characteristics. We are assuming that the WMN BH network would
contain three types of nodes:
\begin{itemize}
\item Street level nodes that have limited range due to blocking
\item Roof-top nodes that have better range due to higher placement
  and that can provide BH connectivity to street level nodes
\item Gateway nodes that have the best visibility to other nodes as
  they are, e.g., co-located with macro cell base stations
\end{itemize}

The core of the topology generator is a simple perturbed grid
generator which is used to create nodes for each of the three
layers. The layers are differentiated by defining different grid sizes
and height ranges for each layer. In practise, the street level has
smallest grid cell size that results in largest number of nodes while there
will be only a few GW layer nodes. The grid cell size of the intermediate
roof-top level is something between the two other layers. After the
node locations are generated, each node pair is evaluated if there is
a link between them. This depends on the distance of the nodes as well
as their z-coordinate that are used to calculate a probability of LOS
between the nodes. A uniformly distributed random value is then drawn
to decide if the link exists or not.

\begin{figure}[!t]
\centering
\includegraphics[width=75mm]{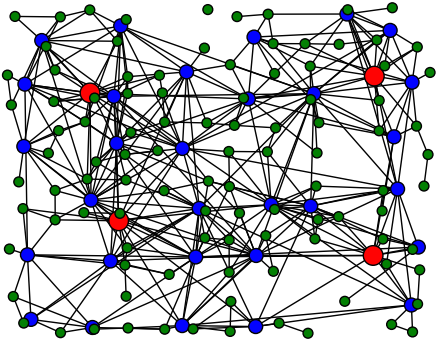}
\caption{An example of dense urban network topology generated by using our model --- the red nodes denote GW nodes, the blue
ones roof-top nodes, and the green ones street-level nodes.}
\label{fig:topo}
\end{figure}

For the following evaluations, we have a set of 16 networks with
number of nodes ranging 66--70 and number of links 284-504. All
networks have 4 GW nodes.





\subsection{Runtimes}

\begin{figure}[!th]
\centering
\includegraphics[width=75mm]{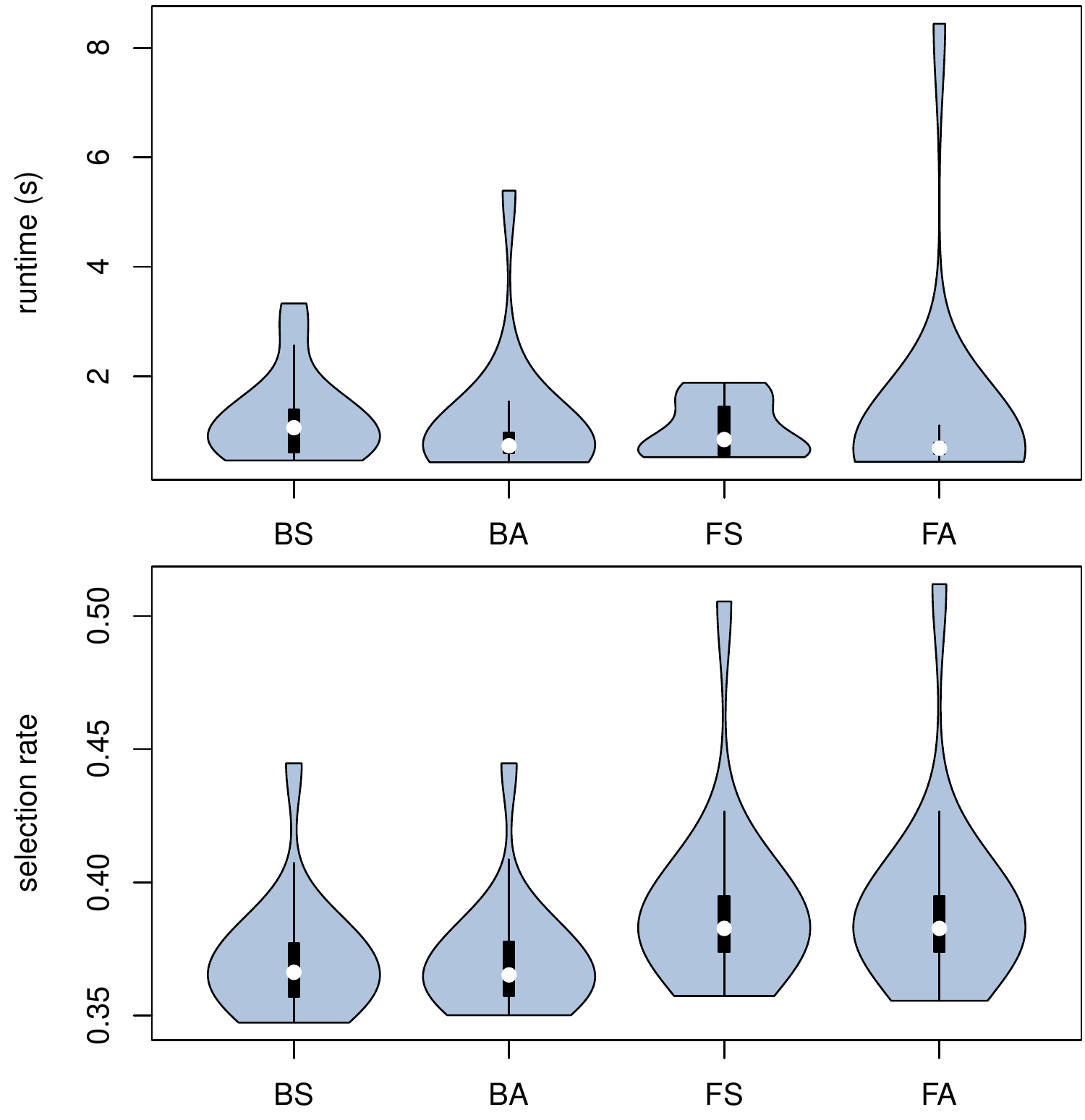}
\caption{Distribution of runtimes (seconds) using different
  strategies (top panel) and distributions of the ratio of selected links (bottom panel).}
\label{fig:runt}
\end{figure}

The runtimes of generation of new network configuration were evaluated
in ``Mid 2015'' Apple MacBook Pro with 4-core 2.8~GHz Intel Core i7 CPU and
by using PyPy 6.0.0. The runtimes were measured by utilizing
\texttt{getrusage()} function from \texttt{resource} library. The
distribution of runtimes is shown in figure~\ref{fig:runt} (top panel). The
overall observation is that on the average the runtimes are relatively low and thus a
new network configuration can be computed whenever needed. When
different strategies are compared, it seems like that having minimal constraints (FS) leads to consistent runtimes. However, the longest runtimes are caused by one network topology and we will investigate if the algorithm could be improved to handle such cases better.

\subsection{Link selection and feedback loop}

The distributions of the ratio of selected links (``links in selected
topology''/''links in original topology'') are shown in
figure~\ref{fig:runt} (bottom panel). While bipartite graph topologies lead in to
shorter runtimes, it seems like it the cost is that the resulting
network has clearly less links compared to free form network.

\begin{figure}[!th]
\centering
\includegraphics[width=80mm]{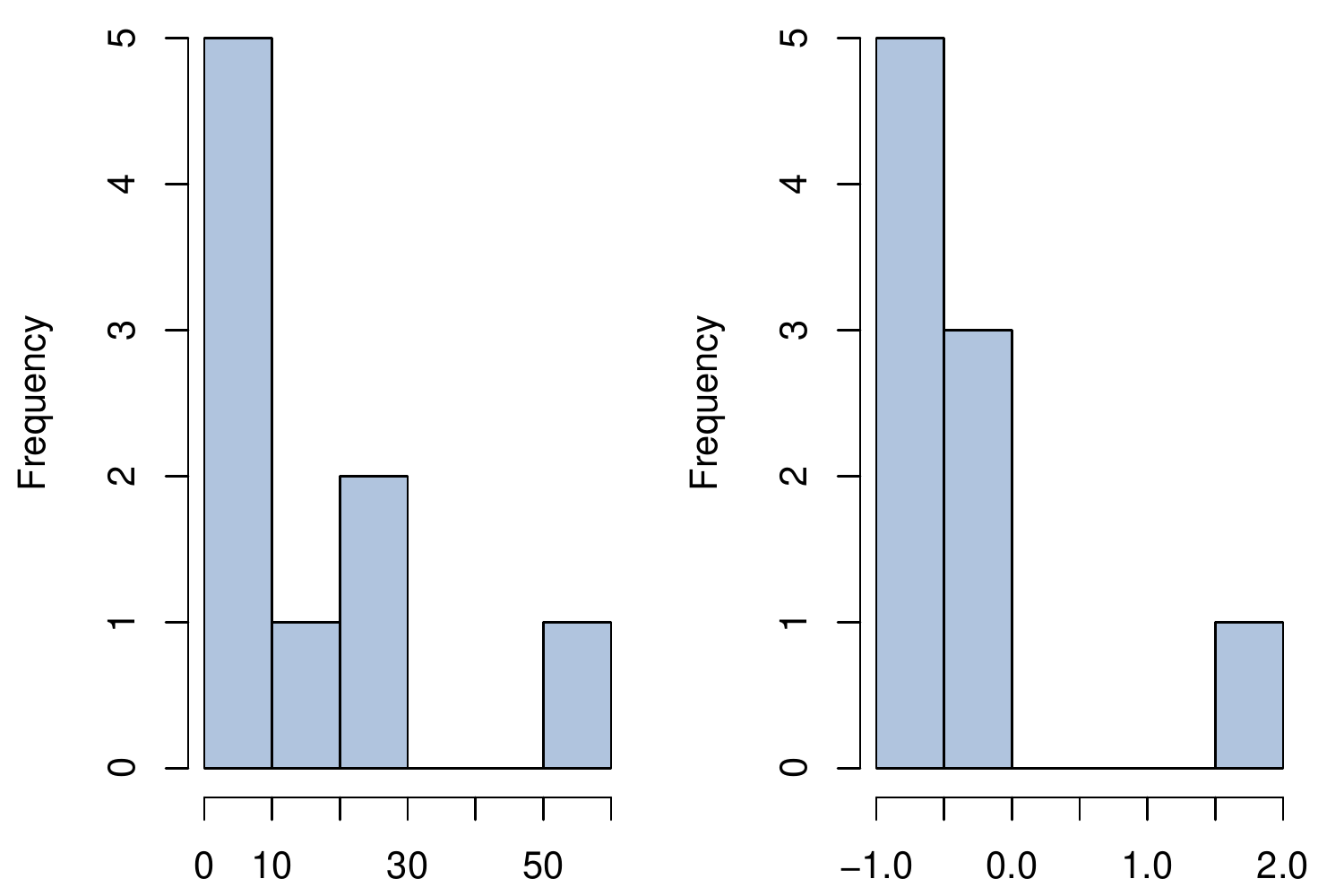}
\caption{Histograms for the size of the ``links to avoid'' list and reduction in
  the number of selected active links after feedback loop activation.}
\label{fig:feedb}
\end{figure}

The feedback loop was activated 4 times in 64 test runs (4 strategy
combinations $\times$ 16 sample networks) and most often one round of
feedback loop was sufficient. While the size of the
``links to avoid'' set was up to 56, the count of selected links
reduced minimally after new rounds of active link
selection. Histograms of ``links to avoid'' set size and reduction in
selected link count are shown in figure~\ref{fig:feedb}. The results
seem to indicate that the proposed feedback loop is very efficient in
removing conflicting links from the topology and, at the same time,
finding new alternative links to avoid reduction in connectivity.

\subsection{Path delays}

\begin{figure}[!th]
\centering
\includegraphics[width=80mm]{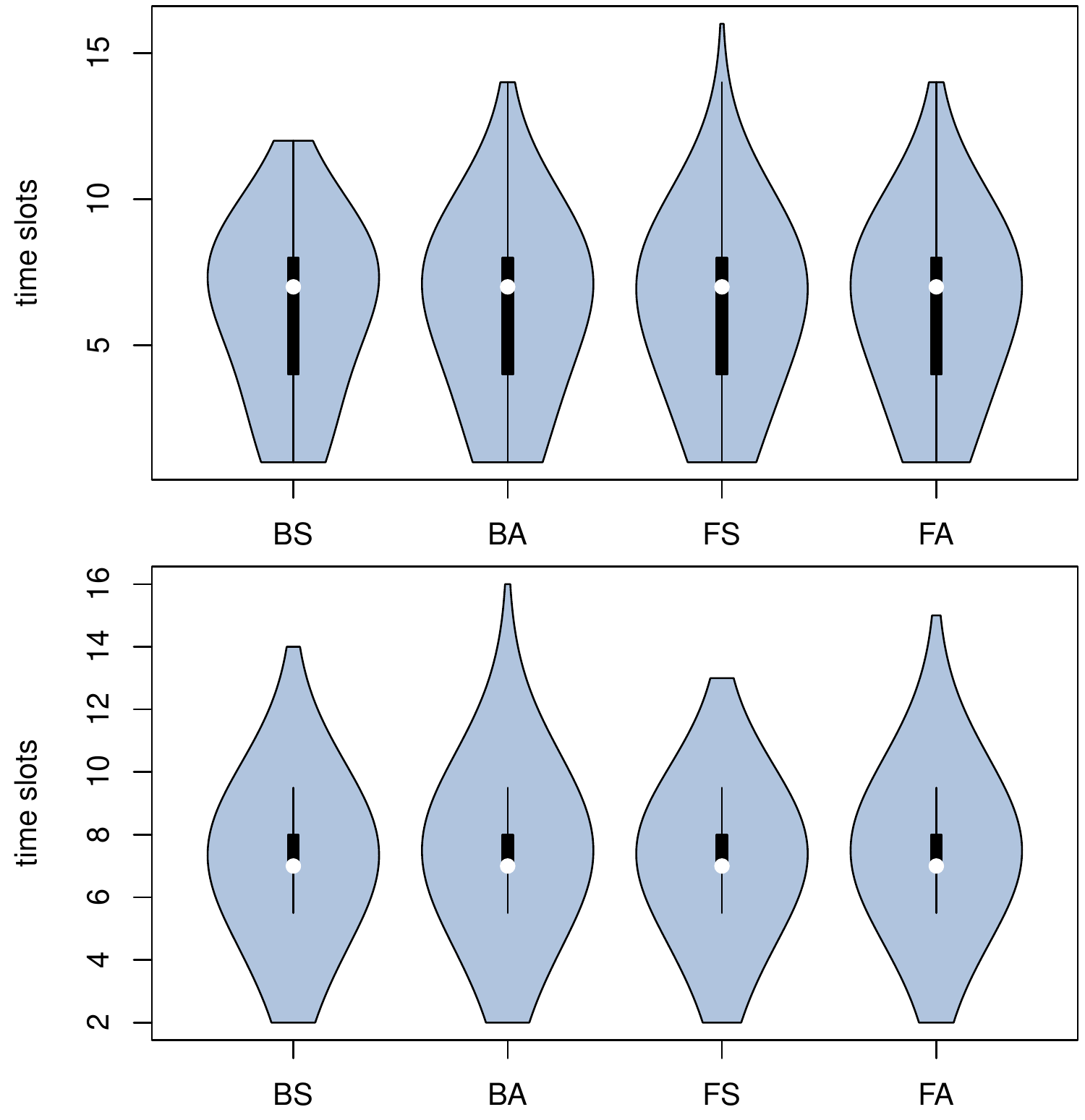}
\caption{Distributions of primary path delays in time slots for different
  strategies. Downstream paths top panel, upstream paths bottom panel}
\label{fig:ppaths}
\end{figure}

The primary path delays for different strategies in both downstream
and upstream directions are shown in figure~\ref{fig:ppaths}. There
seems to be very few differences between different strategy
combinations. This would indicate that while it should be easier to
find an optimal link schedule for a bipartite graph topology, it does
not matter when all the factors are combined. Thus, the selection of
the best strategy should be based on, e.g., reliability of the
resulting network configuration.

\begin{figure}[!th]
\centering
\includegraphics[width=80mm]{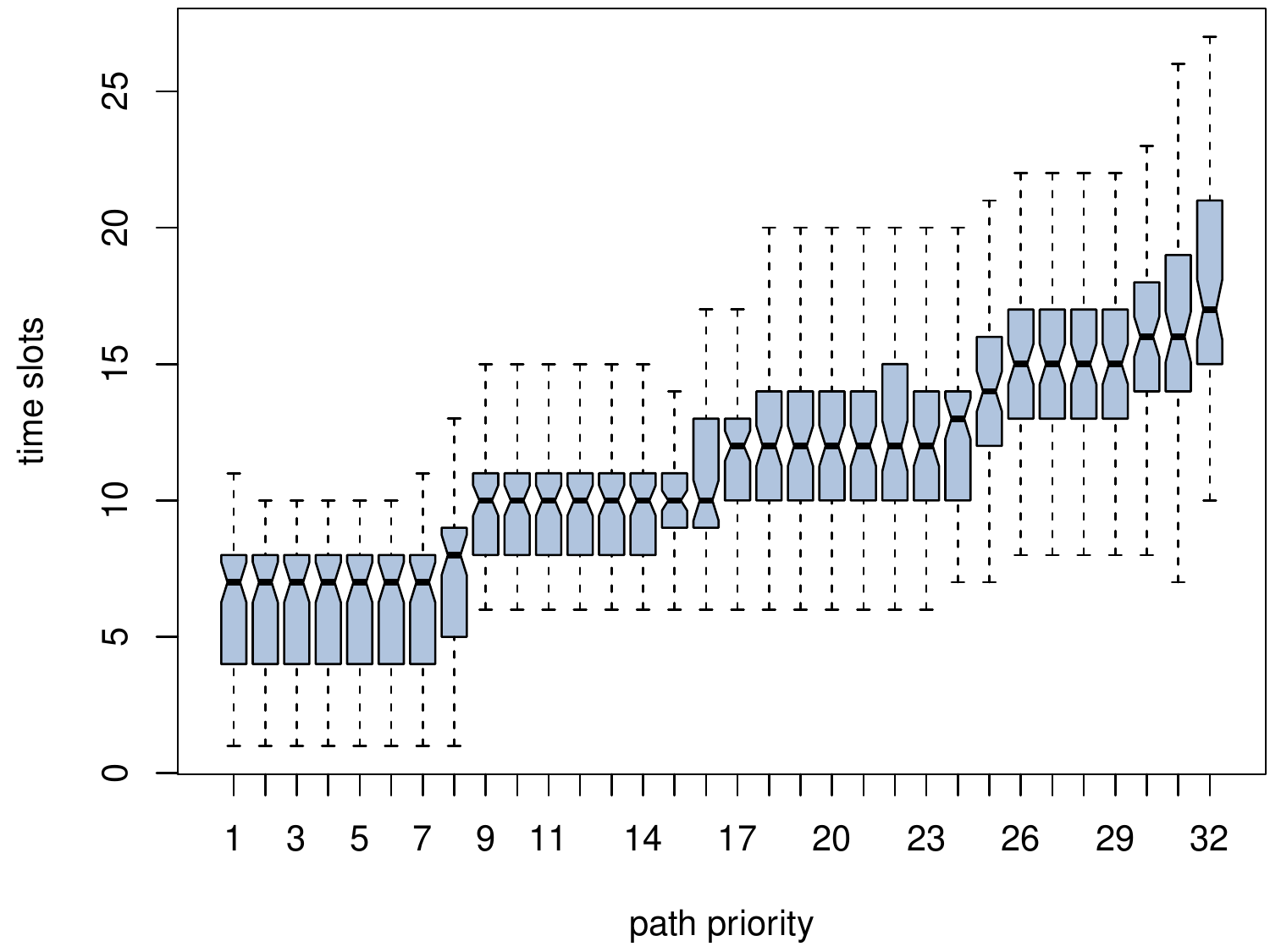}
\caption{Distributions of downstream delays for all alternative paths
  for network 13 using ``bipartite'' and ``allocate all'' strategies.}
\label{fig:p13}
\end{figure}

The distribution of the downstream delays for one network using
``bipartite'' and ``allocate all links at once'' strategy is shown in
figure~\ref{fig:p13}. This distribution indicates that the process
produces a large number of alternative paths that provide low
delays. Thus, paths can changed due to faults or congestion with
minimal effect on transport delays.

\begin{figure}[!th]
\centering
\includegraphics[width=80mm]{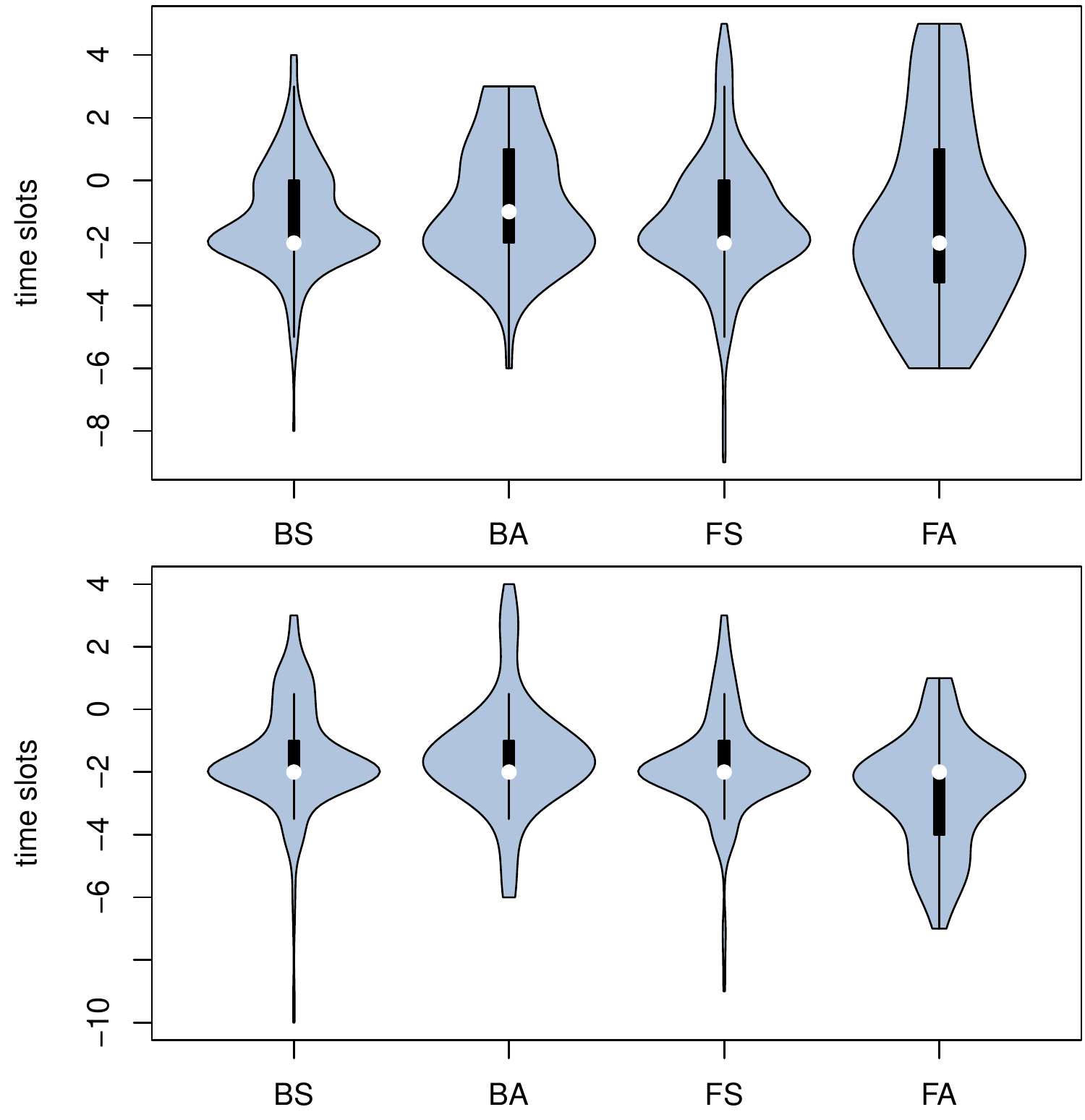}
\caption{Effect of feedback loop in primary path delays: downstream in top panel, upstream in bottom panel.}
\label{fig:ddel}
\end{figure}

The introduction of feedback loop should have positive effect on path delays. However, the nodes with rejected links could suffer due to potentially longer paths to GW. In figure~\ref{fig:ddel}, the distributions of changes in delays of primary paths are shown. The comparison is between optimized schedules before and after feedback loop. As it can be seen, the shorter link schedule resulting from feedback provides delays that are on the average two time slots shorter. In few cases, the delays have increased up to four time slots but this can be allowed as the worst case delay went down in all cases by 1--2 time slots.

\subsection{Disjoint paths}

\begin{figure}[!th]
\centering
\includegraphics[width=80mm]{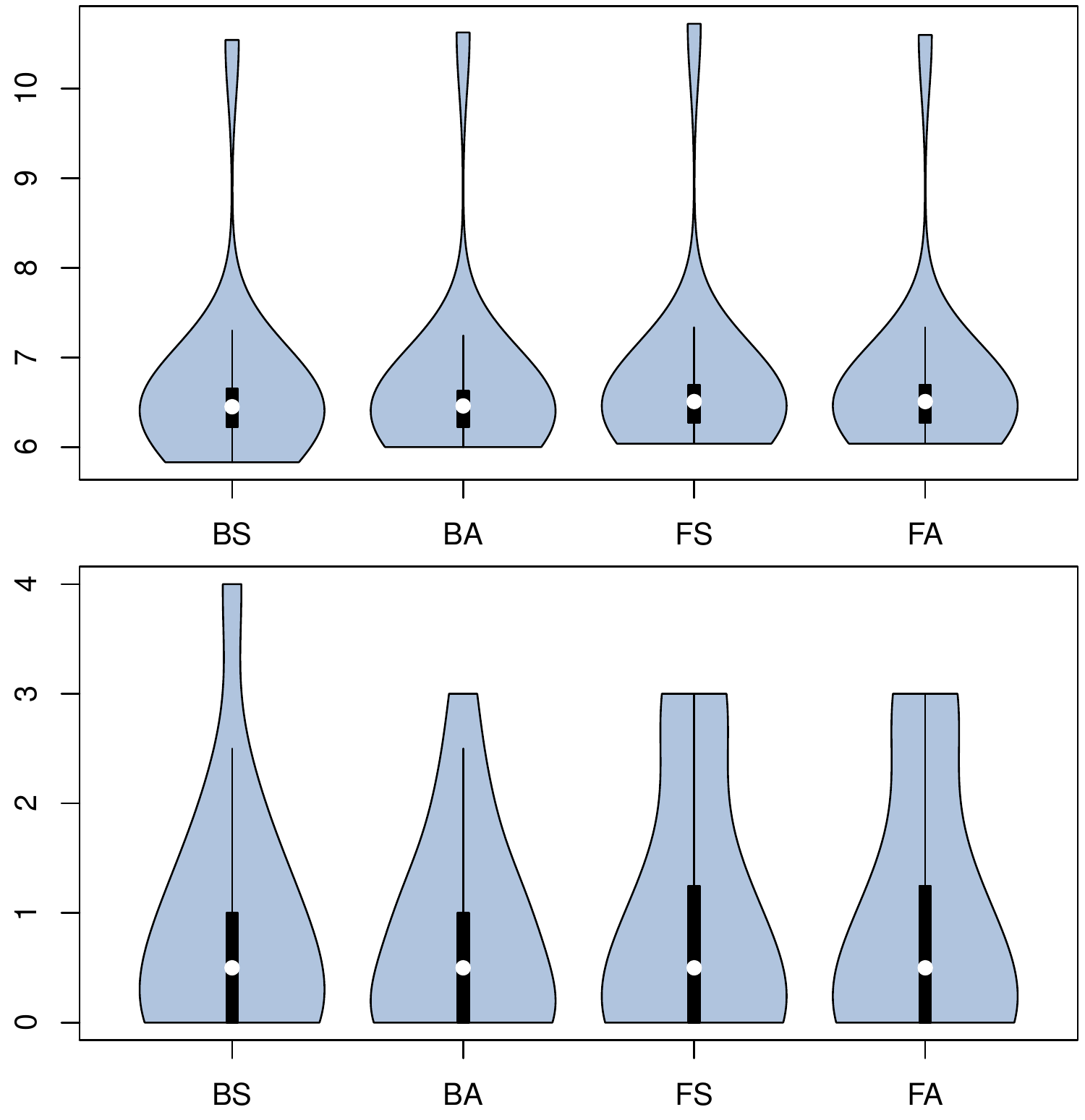}
\caption{Box plot of the distributions of the average number of node disjoint paths
  (top) and the number of nodes without node disjoint paths (bottom) using different
  strategies.}
\label{fig:djo}
\end{figure}

The distributions of node disjoint paths and nodes without disjoint
paths are shown in the figure~\ref{fig:djo}. The distributions seem to be more or less identical which indicates that in this case bipartite network topology would not impose any penalties on reliability.

\subsection{Capacity}

\begin{figure}[!th]
\centering
\includegraphics[width=80mm]{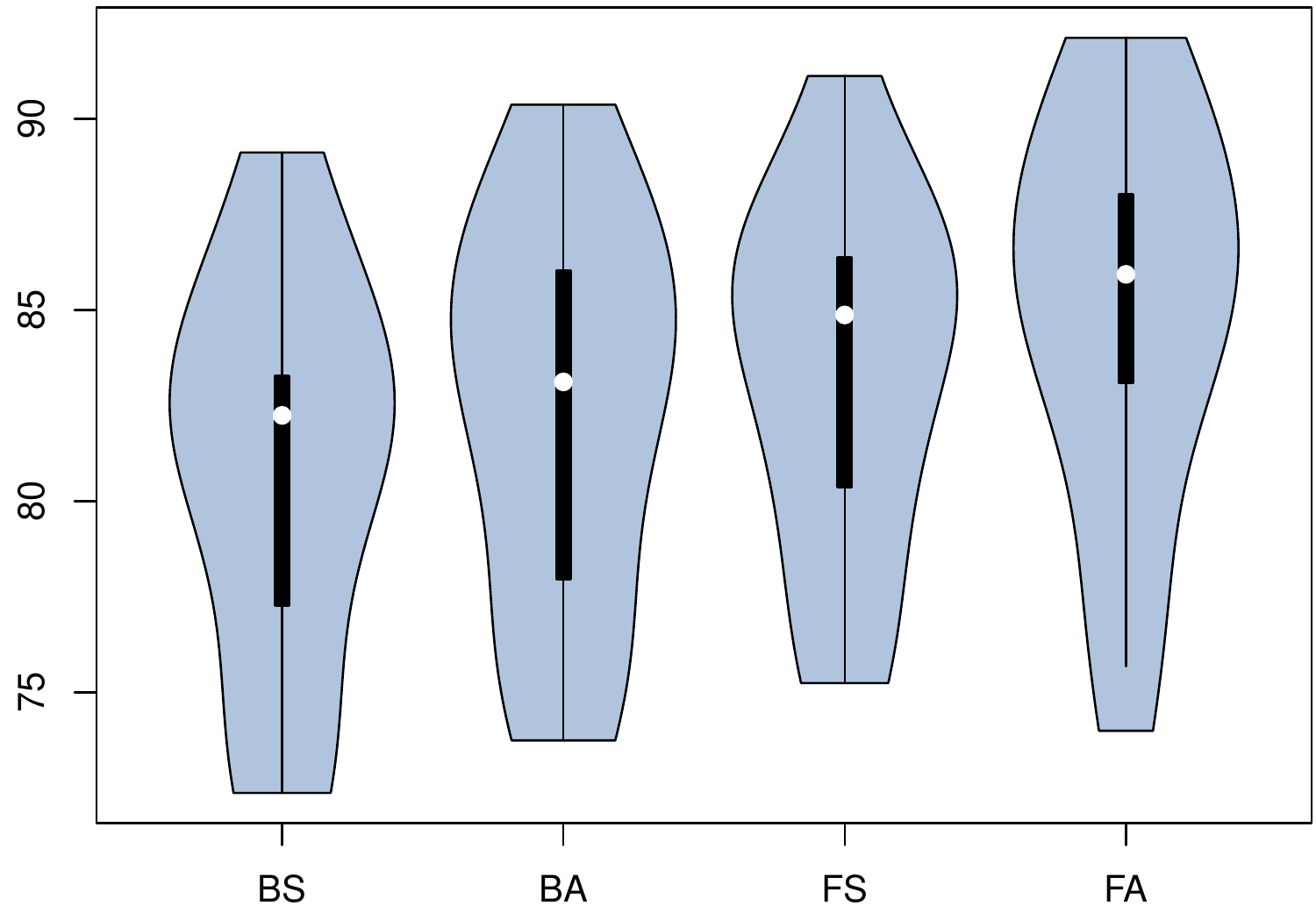}
\caption{Distribution of the average number of active links in one
  time slot.}
\label{fig:alinks}
\end{figure}

The total network capacity is related to the number of active links in
each time slot. The results for different strategies are shown in
figure~\ref{fig:alinks}. It seems like that the number of active links
is closely related to the ratio of selected links
(figure~\ref{fig:runt}) and thus the bipartite graph topology does not
provide any advance over free form network. It seems that the feedback
loop is so efficient in removing ``hard to schedule'' links that the
remaining odd cycles in free form networks do not cause any noticeable
inefficiencies in link schedules. Thus, the conclusion seems to be
that forcing bipartite graph topology does not provide any real
benefits but results in reduced total network capacity. However, the
real throughput is not  necessarily 1:1 related to the number of active
links and dynamic simulations would be needed to find out the exact figures.

\section{Summary}
\label{sec:conc}

A new process to compute routing topology and link schedule for SWMN
is presented in this paper. The main contributions to existing
algorithms is a new active link selection algorithm and a feedback
loop from link scheduling to that phase. Together these features allow
for selecting a suitable subset of links that are used to transport
traffic. The results of evaluation runs with a set of randomly
generated networks indicate that avoiding ``hard to schedule'' links
via feedback loop does not reduce the number of selected links
noticeably.

The other modification to existing algorithms is to add support for
nodes with multiple RUs with a constraint that all RUs at a node can
either transmit or receive at the same time. When the algorithms were
modified for this, it was thought that free form networks with odd
cycles could cause problems for finding an optimal link
schedule. Networks with bipartite graph topology should be easier to
schedule and thus an optional constraint was added to active link
selection algorithm. However, the evaluation results show quite
clearly that free form networks have much better performance which
emphasize the efficiency the feedback loop.

The overall results did show that the original target of minimizing the
worst case delay and proving multiple almost equal cost paths can be
achieved with these new and modified algorithms.



\bibliographystyle{IEEEtran}
\bibliography{IEEEabrv,../bib/wmn.bib}

\end{document}